\documentclass[pre,twocolumn,aps,superscriptaddress]{revtex4}

\usepackage{graphicx}
\usepackage{amssymb}
\usepackage{mathrsfs}
\usepackage[english]{babel}
\usepackage{amstext}
\usepackage{amsmath}

\begin{document}

\title{Parameter-free predictions of the viscoelastic response of glassy polymers from non-affine lattice dynamics}

\author{Vladimir V. Palyulin}
\affiliation{Department of Chemical Engineering and Biotechnology, University of Cambridge, Cambridge, CB3 0AS, United Kingdom}
\author{Christopher Ness}
\affiliation{Department of Chemical Engineering and Biotechnology, University of Cambridge, Cambridge, CB3 0AS, United Kingdom}
\author{Rico Milkus}
\affiliation{Department of Chemical Engineering and Biotechnology, University of Cambridge, Cambridge, CB3 0AS, United Kingdom}

\author{Robert M. Elder}
\affiliation{Polymers Branch, U.S. Army Research Laboratory, Aberdeen Proving Ground, Maryland 21005, USA}
\affiliation{Bennett Aerospace, Inc., Cary, North Carolina 27511, USA}
\author{Timothy W. Sirk}
\affiliation{Polymers Branch, U.S. Army Research Laboratory, Aberdeen Proving Ground, Maryland 21005, USA}

\author{Alessio Zaccone}
\affiliation{Department of Chemical Engineering and Biotechnology, University of Cambridge, Cambridge, CB3 0AS, United Kingdom}

\date{\today}

\begin{abstract}
We study the viscoelastic response of amorphous polymers using theory and simulations. By accounting for internal stresses and considering instantaneous normal modes (INMs) within athermal non-affine theory, we make parameter-free predictions of the dynamic viscoelastic moduli obtained in coarse-grained simulations of polymer glasses at non-zero temperatures. The theoretical results show very good correspondence with rheology data collected from molecular dynamics simulations over five orders of magnitude in frequency, with some instabilities that accumulate in the low-frequency part on approach to the glass transition. These results provide evidence that the mechanical glass transition itself is continuous and thus represents a crossover rather than a true phase transition. The relatively sharp drop of the low-frequency storage modulus across the glass transition temperature can be explained mechanistically within the proposed theory: the proliferation of low-eigenfrequency vibrational excitations (boson peak and nearly-zero energy excitations) is directly responsible for the rapid growth of a negative non-affine contribution to the storage modulus.
\end{abstract}

\maketitle

\section{Introduction}

\begin{figure}\center
\includegraphics[width=7cm]{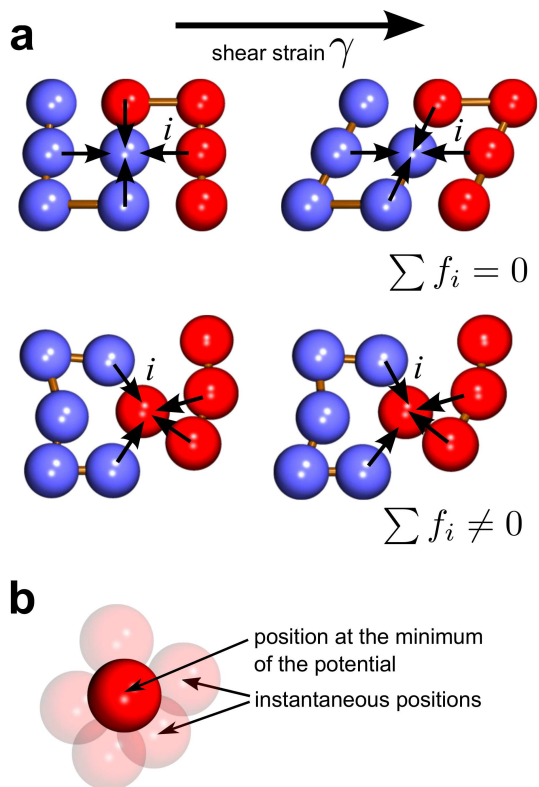}
\caption{(a) Affine (top) vs non-affine (bottom) deformation. In the case of affine deformation the forces acting on a selected atom from its neighbours add up to zero. This changes in the case of non-affine deformation. The net force (affine force field) is non-zero and causes an additional displacement. The blue and red colors denote the atoms belonging to different polymer chains. (b) Instantaneous mode method uses the instantaneous positions of atoms (translucent spheres) rather then the positions at the minimum (solid sphere).}
\label{sketch}
\end{figure}

In centrosymmetric idealised crystals at low temperature, lattice inversion symmetry ensures that atoms are displaced homogeneously under applied deformation, i.e. all displacements are \textit{affine}.
The sum of forces on each atom in the deformed configuration is thus zero thanks to centrosymmetry, leading to straightforward determination of the elastic properties~\cite{BornHuang}. 
Amorphous solids, in contrast, lack such symmetry, and rather exhibit a static snapshot configuration closer to that of liquids. 
As a consequence, net remnant forces under shear displace the atoms from their affine positions, causing the so-called \textit{non-affine} deformation (Fig. \ref{sketch}).
Except at infinite frequency of applied shear, where this non-affine relaxation is inhibited, classical affine microscopic theory fails to predict the mechanical deformation behaviour of amorphous solids~\cite{Lemaitre06}.

Recent works address this shortcoming using theoretical models based principally on solutions of the equation of motion for the non-affine displacement of a tagged atom~\cite{Lemaitre06,PRB2011,AlessioEugene13,tanguyPRB2017}. A correction to the stress response can be obtained~\cite{Lemaitre06} by enforcing mechanical equilibrium on every atom at all steps in the deformation: the forces arising when nearest-neighbour atoms try to find their affine position are relaxed at all steps, and the displacement field which satisfies the mechanical equilibrium contains additional \emph{non-affine} displacements on top of the affine ones. This non-affine deformation framework is crucially dependent on the vibrational density of states (VDOS), since one needs to evaluate this force-relaxation over the whole space of degrees of freedom, and on a quantity which describes how the force field due to affine displacements depends on eigenfrequency~\cite{Lemaitre06,Rico17}.
Both the VDOS and the eigenmode-correlator of the affine force-field are found by diagonalisation of the Hessian matrix of the system. Importantly, the low-frequency part of the VDOS makes the most substantial contribution to the viscoelastic moduli, which is consistent with the observation of anomalous soft modes in glasses \cite{Parisi2015,Xu,DeGiuli2014}. These modes arise from the lowest energy barriers for rearrangements of atoms \cite{wyart2009,Manning2011,procaccia2012}.

The non-affine deformation framework as described above assumes that the system is athermal. Practically, this means that the system resides at, or very close to, a local minimum of the potential energy, fluctuations are negligible, and features of the behaviour are described by standard normal mode analysis.
In reality, however, the system spends most of its time off the energy minimum even at low non-zero temperatures.
One approach to the computation of the VDOS in this case is based on instantaneous normal modes (INM)~\cite{Stratt1995,INMbook}. Instead of using the energy of the system at the potential minimum as in conventional normal mode analysis, single snapshots of the system are considered and averaging is performed over the snapshots. This method was earlier applied to liquids \cite{Keyes1997} as well as glassy systems such as LJ glass \cite{Bembenek96}, silica \cite{Bembenek2001} and proteins \cite{protein2009}. The instantaneous normal modes of liquids have predictive power only at high frequencies \cite{gezelter1997,stratt1997,stratt1998}. For amorphous solids these modes can be defined at much slower timescales, i.e. some properties of solids as well as the glass transition can be associated with INMs \cite{bembenek1995}.

In this work we show that a combination of the non-affine theory and the INM approach produces quantitative parameter-free predictions for viscoelastic behavior of amorphous solids. We compute the VDOS and correlators of affine forces from averages over snapshots of the system rather than from a long-time average, which is necessary for taking the temperature-dependent unstable modes into account~\cite{INMbook}. As a model system we use polymer glasses. However, the method and the results are relevant far beyond this particular system. Our choice is based on high practical relevance of polymer materials and the fact that their complexity better highlights the usefulness of the theory. 
The storage modulus $G'$ computed within our framework matches results from simulations across a wide range of temperatures, the only departure being at low frequencies around and above $T_g$, where the timescale for structural relaxation competes with that of the externally applied shear, thus violating our harmonic approximation. The loss modulus $G''$ is qualitatively predicted across the full range of temperatures explored, though the quantitative agreement is not as good as that for $G'$.

As temperature is increased, the system spends more time further away from local potential energy minima~\cite{Goldstein69}, leading to increased importance of INMs as well as local internal stresses (because away from the minimum the first derivative of intermolecular interaction is non-zero). Once the internal stresses are included, the analysis produces some negative eigenvalues, which correspond to imaginary frequencies in the VDOS and therefore to non-propagating relaxation modes down from saddles (these modes are localised). The number and the density of these relaxation modes grow with temperature, reflecting an increasing instability of the system. Importantly, this growth is continuous across the glass transition, providing evidence that the glass transition, at least in its mechanical manifestations, has hallmarks of a crossover rather than a true phase transition. As such, our approach offers fundamental insights into the mechanics of amorphous materials across the glass transition, as well as a robust prediction of the viscoelastic properties of real amorphous materials across both solid and liquid states.

\section{Theory and Simulations}

We focus on the case of small deformations, within the regime of linear viscoelasticity and avoiding complications such as shear banding~\cite{Parisi2017}, local anharmonicity and nonlinear plastic modes~\cite{Lerner2016,BouchbinderPNAS17}, all of which lie out of the scope of our approach. Our formalism below is written accordingly.
The non-affine theory \cite{Lemaitre06,PRB2011,AlessioEugene13} computes corrections to the elastic moduli due to additional displacements caused by an extra net force from neighbours in the case of non-centrosymmetric materials (Fig. \ref{sketch}).
The non-affine displacements cause softening of the material.  The corresponding correction to the elastic free energy is negative and for shear deformation it can be expressed as

\begin{equation}
F=F_\mathrm{A}-F_\mathrm{NA}=F_\mathrm{A}-\frac{1}{2}\sum_i\frac{\partial\mathbf{f}_i}{\partial\gamma}\frac{\partial \mathbf{r}_i}{\partial\gamma}\gamma^2,
\label{FreeEnergy1}
\end{equation}
where the affine part $F_\mathrm{A}$ was given already by Born and Huang \cite{BornHuang}, $\gamma$ is a shear angle (shear strain amplitude), 
$\mathbf{f}_i$ is the net force which acts on the atom $i$ in its affine position (see Fig. 1), $\mathbf{r}_i$ is radius-vector to $i$-th atom, $-F_\mathrm{NA}$ is the non-affine contribution to free energy. If the interactions between the atoms are described by harmonic central forces then Eq. \ref{FreeEnergy1} can be written as~\cite{lutsko89}

\begin{figure*}\center
\includegraphics[width=15.1cm]{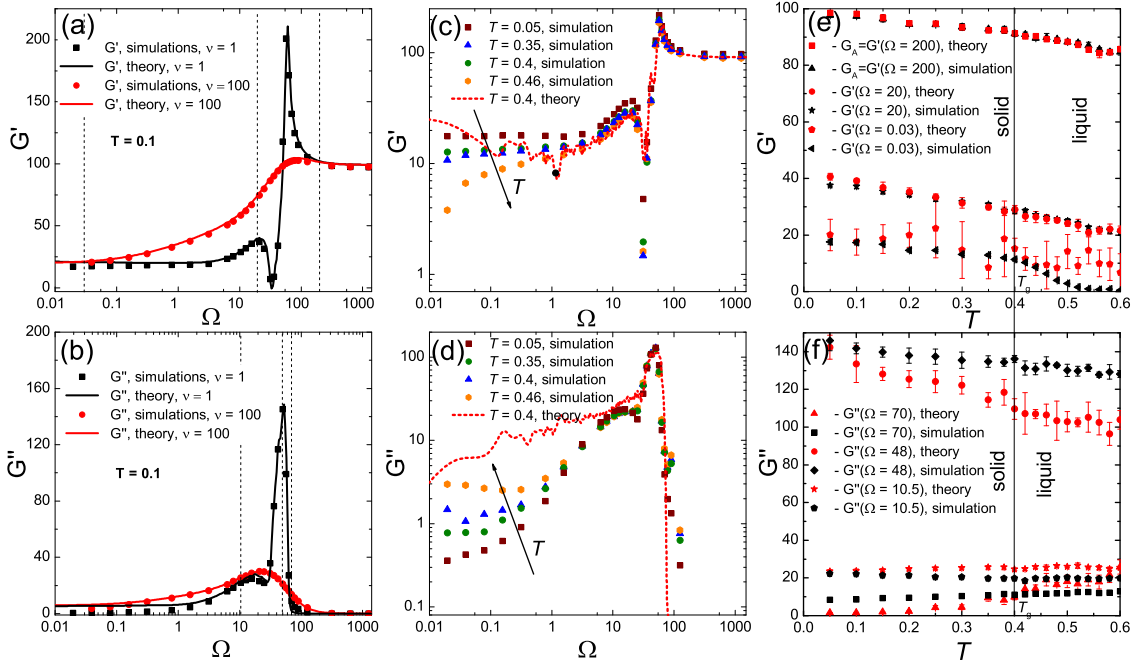}
\caption{Dependence of viscoelastic moduli on the frequency of external shear $\Omega$ and temperature $T$ obtained from theory and simulations. (a),(b) respectively, $G'(\Omega)$ and $G''(\Omega)$ for two different values of friction, which correspond to $\nu$ in theory and connected to a Langevin thermostat damping as $\nu=m/\xi$. $T=0.1, N=100, M=100$. The dashed vertical lines correspond to the frequencies in Figs. 2(e),(f).  (c),(d) respectively, $G'(\Omega)$ and $G''(\Omega)$ from LAMMPS simulations for the dependence of storage modulus $G'$ as a function of the deformation frequency for different temperatures, $N=50, M=100$. To highlight the applicability of the theory the red dashed line shows the result for $T=0.4$. (e) Comparison of theoretical predictions with the mechanical spectroscopy simulations for $G'$ as a function of temperature. The data are shown for low ($\Omega=0.03$), middle ($\Omega=20$) and high frequencies ($\Omega=200$), $N=50, M=100,\nu=1$. (f) Comparison of theoretical predictions with the mechanical spectroscopy simulations for $G''$ as a function of temperature. The data are shown for three frequencies in the vicinity of the loss peak, ($\Omega=10.5,\Omega=48,\Omega=70$), $N=50, M=100,\nu=1$.  In all cases the theoretical and simulation results were obtained for 5 realisations of a corresponding system and then averaged.}
\label{Panel1}
\end{figure*}

\begin{equation}
F=F_\mathrm{A}-\frac{1}{2}\mathbf{\Xi}_i\mathbf{H}^{-1}_{ij}\mathbf{\Xi}_j\gamma^2,
\end{equation}
where an affine force field $\mathbf{\Xi}_i$ is responsible for a force $\mathbf{f}_i=\mathbf{\Xi}_i\gamma$ acting on atom $i$, and $\mathbf{H}_{ij}$ is the Hessian (which for a system of $N$ particles has size $3N\times3N$), which includes the internal stresses induced by thermal fluctuations via the INMs (for details see SI). Summation over repeated indices is implied.
Assuming the system dwells in the vicinity of a local energy minimum, the microscopic equation of motion for a particle of mass $m$ can be written as that of a damped harmonic oscillator

\begin{equation}
m\ddot{\mathbf{r}}_{i}+\nu\dot{\mathbf{r}}_{i}+\mathbf{H}_{ij}\mathbf{r}_{j}=\mathbf{\Xi}_{i}\gamma,
\label{oscillator}
\end{equation}
where the second term characterises energy dissipation due to viscous damping (with friction coefficient $\nu$), the third term is the harmonic force pulling the particle back into the minimum and the right hand side is the non-affine force, which depends on the shear amplitude $\gamma$. 

Transformation of Eq. (\ref{oscillator}) into Fourier space allows us to obtain an expression for the complex viscoelastic modulus, which in the continuous limit reads (see \cite{Lemaitre06} for  more details) 
\begin{equation}
G^{*}(\Omega)=G'+iG''=G_A-\frac{1}{V}\sum_k\frac{\Gamma(\omega_k)}{m\omega^2_k-m\Omega^2+i\Omega\nu},
\label{CModulus}
\end{equation}
where $\Omega$ is the applied strain frequency, $\Gamma(\omega_k)=\langle\mathbf{\Xi}_{p}^2\rangle_{\omega_p\in[\omega,\omega+d\omega]}$ is the affine force field correlator, and $G_\text{A}$ is the affine contribution~\cite{BornHuang}. The important quantity, which implicitly enters the expression, is the vibrational density of states $D(\omega_k)$ normalised as $\sum_k D(\omega_k)=1$. The inputs to the expression for $G(\Omega)$ are the eigenvalues ($\omega_k$) and eigenvectors (through $\Gamma(\omega_k))$ of the Hessian matrix $\mathbf{H}_{ij}$, which we obtain from molecular simulations of glassy polymer configurations as described in the following. It is important to stress here the difference to Ref. \cite{Lemaitre06}. In Ref. \cite{Lemaitre06} it is assumed that the Hessian is positive definite. In our approach consideration of INMs leads to the appearance of the negative eigenvalues of the Hessian, i.e. summation includes the imaginary part of the frequency (VDOS) spectrum. 

In the thermodynamic limit the summation can be expressed as an integral over the frequency domain: 

\begin{equation}
G(\Omega)=G_A-\frac{3N}{V}\int_{C}\frac{D(\omega)\Gamma(\omega)}{m\omega^2-m\Omega^2+i\Omega\nu}d\omega,
\end{equation}
where $\Omega$ is as an applied strain frequency, $G_A$ is derived in Appendix A and $\nu$ is a constant friction. The integration should be performed over a contour which includes the positive part of imaginary axis of $\omega$ values and the positive part of real axis.

Amorphous polymeric systems of $N=50$ and 100 linear homopolymer chains with polymerisation degree $M=100$ were modelled in a periodic domain using LAMMPS~\cite{LAMMPS}. We explored the role of chain length on the vibrational properties in an earlier article and verified therein that there is no size dependence in this system when $N\geq50$~\cite{macro2018}. We adopt the conventional Kremer-Grest bead-spring model~\cite{KGmodel}, i.e. polymer backbone covalent bonds were simulated using a finite extensible nonlinear elastic (FENE) potential, while non-bonded interactions were represented by a shifted Lennard-Jones (LJ) pair potential. For the FENE potential $U_\mathrm{FENE}=-0.5KR_0^2\ln\left[1-\left(\frac{r}{R_0}\right)^2\right]$ the parameters were set as $K=30,R_0=1.5$.
Our motivation for this choice of $K$ is threefold: i) it is consistent with most of the literature on bead-spring polymeric simulations; ii) it is large enough so that we are close to the limit of stiff chains, where backbones cannot be stretched; iii) it provides a separation in characteristic frequency that is instructive when interpreting the VDOS.
For LJ potential $U_\mathrm{LJ}=4\varepsilon\left[\left(\frac{\sigma}{r}\right)^{12}-\left(\frac{\sigma}{r}\right)^6-\left(\left(\frac{\sigma}{r_c}\right)^{12}-\left(\frac{\sigma}{r_c}\right)^6\right)\right]$ the constants were chosen to be $\varepsilon=1,\sigma=1$ and the cutoff radius $r_c=2.5$ (which matches that used in the computation of $\mathbf{H}_{ij}$). Bead trajectories are updated according to Langevin dynamics, with a damping constant $\xi$ (which is related to the theoretical damping term by $\xi = m/\nu$). $\varepsilon$ sets the LJ energy scale and $K$ is the bond energy scale, where $K/\varepsilon = 30$. With reference to fundamental units of mass $M$, length $d$, and energy $\mathcal{E}$, we set $\sigma=1$ and $m=1$, giving a time unit of $\tau=\sqrt{m\sigma^2/\varepsilon}$. We report frequency in units of $1/\tau$ and temperature in units of $k_bT$ throughout. We equilibrate the system in a melted state at $T= 1.0$, maintaining zero external pressure using a Nose-Hoover barostat. We then cool the system by decreasing $T^*$ with a characteristic timescale $\tau_c > 10^5\tau$ following the equation $T^*(t) = T_\mathrm{start}(1 - t/\tau_c) + T_\mathrm{end}(t/\tau_c)$ until the target temperature is reached. A typical size of a cubic simulation box after the equilibration was about 17 length units for systems with 5000 monomers and about 21 units for 10000 monomer system.

We then obtain the viscoelastic moduli by mechanical spectroscopy, applying small amplitude oscillatory simple shear strain to the sample as in Refs. \cite{tanguyPRB2017,Keblinski}. For every sample we have simulated 20 periods of the applied periodic strain, beyond which there is no further change to the stress-strain relationship (see illustrations in the Appendices). From the stress-strain curves we compute the storage $G'$ and loss $G''$ moduli
\begin{equation}
G'=\frac{\sigma_0}{\epsilon_0}\cos\delta, G''=\frac{\sigma_0}{\epsilon_0}\sin\delta,
\end{equation}
where $\sigma_0$ is the average amplitude of stress, $\epsilon_0$ is the amplitude of strain (fixed at 2\%) and $\delta$ is a phase shift between the two. We have checked that 2\% strain is still in the linear regime and produces the same results for $G'$ and $G''$ as strains of 0.5\% and 1\%. We have encountered the lower limit of simulating the moduli at lower frequencies ($\Omega\lesssim0.001-0.01$) due to the increase of the noisiness of the simulation results and the difficulty of approximation of the data with trigonometric functions.

The cooling procedure is initiated with a random seed so that each realisation has a different arrangement of the beads. In order to estimate the error we have obtained 5 different realisations of the frozen disorder in order to smoothen out the sample-specific effects. The error bars in Figs. 2(e,f) and Fig. 3 (a) are standard deviations of the averaging over these realisations. The theoretical results also have some spread in values and, thus, we show their standard errors. We do not show errors in Figs. 2(a,b,c,d) to avoid obstructing the view. The size of the error bars in all cases can be understood from Figs. 2(e,f).

\section{Viscoelastic moduli}

Using the methods described in the previous section we first compare theoretical predictions of $G'$ and $G''$ (obtained using instantaneous static simulation snapshots to form the VDOS with INM) against simulation results obtained by mechanical spectroscopy for low temperature, $T=0.1$. Fig. 2(a,b) shows the storage (a) and loss (b) moduli for low and high internal friction $\nu$ (cf Eq.~\ref{CModulus} and $\xi$ in simulation description), respectively. For the corresponding friction coefficients, the theory provides a good quantitative prediction of the mechanical spectroscopy data (continuous lines and symbols, respectively, in Fig. 2(a,b)) without any free parameter. 

The peaks in $G'(\Omega)$ (Fig. 2(a)) correspond to the strongest resonances in the system. The highest peak is located close to the resonance frequency of FENE bonds $\omega_{\mathrm{FENE}}\approx31.3$, while the peak at lower frequencies is located between the FENE bond peak and the LJ resonance frequency ($\omega_{\mathrm{LJ}}\approx7.56$). Between the peaks $G'$ falls to very small values. With an increase of the friction, the transition from the high frequency regime to low frequency regime is smoother, with resonances being smeared off. At very low frictions (not shown) the theoretical curves show a number of small single frequency resonances, consistent with low temperature results produced by non-affine theory without INMs for amorphous silica glasses~\cite{tanguyPRB2017}. In our case the friction is constant (Markovian) by construction of the molecular simulations with Langevin thermostat, and not a function of $\Omega$. In some real materials such as metallic glass this may not be the case, and an extension that includes memory-function for the friction has been reported recently~\cite{bingyu}.
In the context of interpretation of experiments one would need to determine the effective friction. The friction could be found for instance from an analysis of the atoms' trajectories under the assumption that the atoms move according to the Langevin equation.

Increasing $T$ mostly affects the values of the moduli at low frequency, shown in Fig. 2 (c,d). The storage modulus drops substantially after the temperature exceeds the glass transition temperature $T_g=0.4$, i.e. upon "disorder-assisted" melting~\cite{AlessioEugene13}. On the contrary, the loss modulus grows with temperature, which also reflects a growth in dissipation defined usually as $G''/G'$. The glass transition temperature $T_g$ of our system, which we determined earlier from the change of thermal expansion coefficient for this system \cite{stiffness}, is consistent with the behaviour of the moduli at low frequencies. The data from simulations for low frequency $\Omega=0.03$ (Fig. 2(e)) show that $G'$ drops significantly at $T\approx0.4$ and thereafter smoothly tends to zero, which is consistent with recent results \cite{baschnagelPRL17,ivan18} obtained from the stress-fluctuation formalism \cite{baschnagelPRE15}. In order to show the limitations of our approach we plot dashed lines with theoretical results in Figs. 2(c),(d) for $T=0.4$. It is clear that at low frequencies the simulation and the theory do not match each other well at the glass transition point. In order to better compare the theory with simulation data at different temperatures we have plotted values of the storage modulus for high frequencies and intermediate frequencies in Fig. 2(e). We see that the theory is able to predict the storage modulus even in the liquid phase for moderate to high oscillation frequencies $\Omega$. Fig. 2(f) shows the comparison of theoretical and simulation data for the loss modulus. Since at very low and very high frequencies the values of $G''$ become very small and rather noisy, we have chosen a set of frequencies different from that used for the plot of the storage modulus (Fig. 2(e)). We observe that the match between the theoretical predictions and the simulation data is less quantitatively accurate, but is qualitatively good at intermediate and high frequencies.

\begin{figure}
\includegraphics[width=8cm]{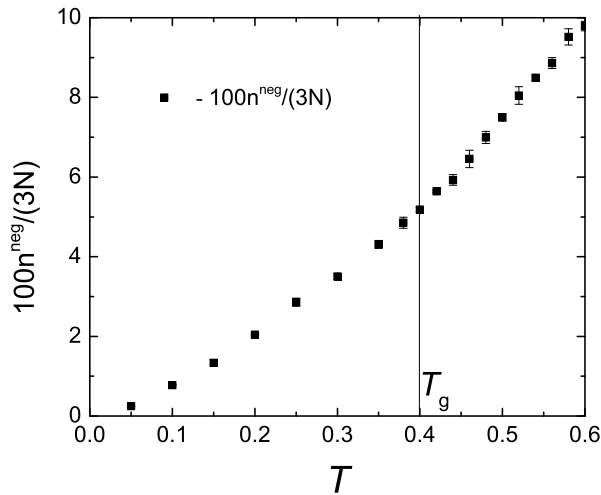}
\caption{Percentage of negative eigenmodes of the Hessian Matrix as a function of temperature, $N=50, M=100$. The data points were obtained by averaging over 5 realisations for every point}
\label{PerNegEig}
\end{figure}

\begin{figure*}
\includegraphics[width=17.8cm]{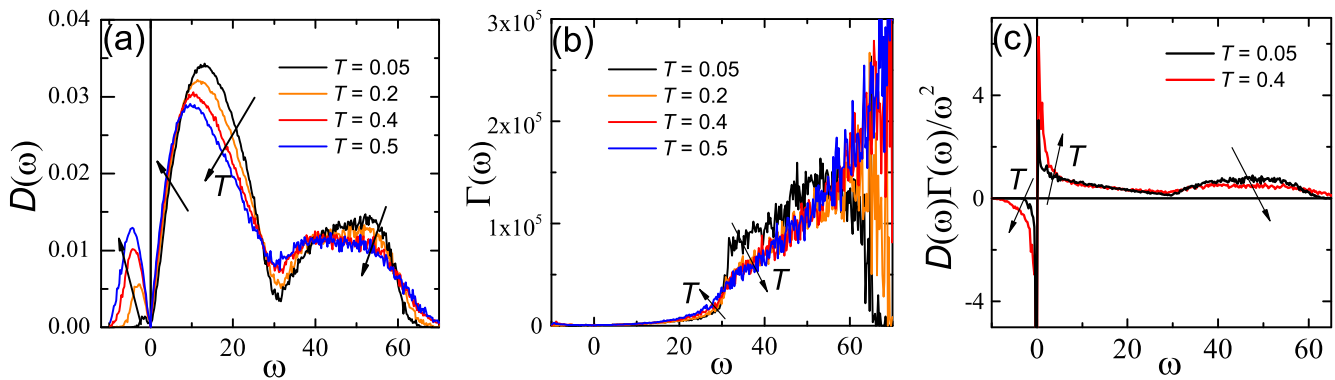}
\caption{Evolution of microscopic characteristics of polymer glasses with temperature. (a) Evolution of theoretically computed VDOS with temperature. $N=50, M=100$. One can see that the boson peak moves towards smaller frequencies. The negative part of the frequency axis shows absolute values of imaginary frequencies, which are obtained from the negative eigenmodes. Arrows show the change with an increase of temperature; (b) Correlator $\Gamma(\omega)$ for the same set of temperatures as in (a) $N=50, M=100$; (c) The quantity $D(\omega)\Gamma(\omega)/\omega^2$ for temperatures $0.05$ and $0.4$ as in (a) $N=50, M=100$.}
\label{Panel2}
\end{figure*}

Although the non-affine approach was developed originally for the athermal case~\cite{lutsko89,Lemaitre06}, the success of its predictions shown in Figs 2(a,b,e,f) suggests that our inclusion of INMs and internal stresses makes it applicable over a broad range of temperatures, even above $T_{g}$. We have checked that upon computing the VDOS in the standard way, taking configurations from long-time energy-minimised simulations (which washes out the relaxation modes and the internal stresses), the comparison with simulation data is much worse.

With an increase of $T$ to a value close or above $T_g$, the predictions of INM non-affine theory get less quantitatively accurate at low frequency (see Fig 2(e)). We believe that a growing mismatch with mechanical spectroscopy data is caused by the increased anharmonicity and by swaps of nearest-neighbours. Even at non-zero temperatures for moderate and high frequencies of shearing, the atoms mostly dwell around their local minima, hence the harmonic approximation holds, although internal stresses resulting from displacements off the minima are important and are taken into account via the INM in our theory. Increase of the external shearing period (for $\Omega\ll 1$) leads to the increase of atom mobility, because the atoms have more time to relax their positions with respect to their neighbours. Indeed, in the SI we show results illustrating that the number of nearest neighbour changes increases with period and temperature. Moreover, we observe a crossover to a more active increase in number of swaps of nearest neighbours once the temperature exceeds $T_g$. Since these swaps are not taken into account in Eq. (\ref{oscillator}), a mismatch arises at the low frequencies once $T$ approaches and even more when it exceeds $T_g$. One can also notice that the agreement between the theory and simulations for the storage modulus $G'$ (Fig. 2(e)) is better than for the loss modulus $G''$ (Fig. 2(f)) except for very low temperatures. The possible reason can be attributed to the increase of the dissipation with temperature which in turn leads to an increased number of monomer relocations.

\section{Analysis of vibrational excitations}

If the Hessian matrix is computed for instantaneous atom positions instead of positions taken from a minimised energy state of the system, then the diagonalisation of $\mathbf{H}_{ij}$ produces negative eigenvalues \cite{Keyes1994,Stratt1995,INMbook}. Their presence indicates that the local slope and curvature of the energy hypersurface has nonconvex components, so the density of these modes is linked to non-propagating relaxation from local saddles (produced by thermal excitation) in the energy landscape ~\cite{Stratt1995}. Since the instantaneous positions deviate more from their rest positions as the temperature is increased, the behaviour of the negative eigenvalues, which correspond to imaginary frequencies, correlates with softening of the material and the transition from solid-like to liquid-like properties.

In Fig. 3 we plot the overall fraction of vibrational modes that are negative as function of temperature. The behavior of unstable modes is rather similar to silica glasses from Ref.~\cite{Bembenek2001} and the small protein system from Ref.~\cite{protein2009}: they appear once the temperature gets above zero. This fraction could also be obtained by the ratio of the areas of imaginary part of VDOS to the real one (see Fig. 3(b)). To understand the nature of these negative modes, we compute a standard measure of localisation, the participation ratio $p(\omega_j)$. For an eigenmode with the eigenfrequency $\omega_j$, it is defined as:
\begin{equation}
p(\omega_j)=\frac{\left(\sum_{i=1}^Nu^2_i(\omega_j)\right)^2}{N\sum_{i=1}^Nu^4_i(\omega_j)},
\end{equation}
where $u_i(\omega_j)$ is the total amplitude of the $i$-th atom's eigenvector. $p(\omega_j)$ quantifies the number of particles participating in a single mode. For instance, for an isolated particle one has $p=1/N$, while if all particles are involved in a single mode  $p=1$. For an amorphous polymer solid, which we consider here, the participation ratio shows substantial change with an increase in temperature (Fig. 5). 

As is customary~\cite{Stratt1995}, we show imaginary frequencies as negative ones. For low temperature ($T=0.05$) almost all frequencies are real and the values of participation ratio for the frequencies close to 0 are in the range 0.05-0.15. As the temperature approaches and crosses $T_g$, more modes become unstable. At around $T_g$ the shoulder of the first band of participation ratio (with values 0.4-0.5) is crossing into the unstable domain with values of participation ratio at $\omega\sim0$ in the range 0.35-0.45. However, there is no special feature visible at $T_g=0.4$ and the process appears smooth across the glass transition. This calculation also shows that the unstable relaxation modes are localised, with low values of participation ratio of imaginary-frequency modes. 

One can notice that even for the data at the lowest temperature $T=0.05$ the participation ratio at low (real) frequencies has values around 0.2-0.4, i.e. the long-wavelength phonons, which are supposed to have $p\approx 0.6$, are not apparent. This does not mean, however, that the phonons are not present. The necessary truncation of the system size in simulations often leads to small values of $p(\omega)$ at the phonon frequencies, which is not unusual for delocalised modes \cite{Mazzacurati1996}. Interestingly, just like in the case of dependence of viscoelastic moduli on the temperature, there are no obvious signatures of a transition in $p(\omega)$ at or near $T_g$.

This corroborates the seminal idea of Y. Frenkel~\cite{Frenkel}, that, in the case of amorphous solids, there is a continuity between the liquid and solid states at least for the mechanical properties, and that, indeed, liquids behave like solid glasses at sufficiently high frequency of deformation.

\begin{figure}\center
\includegraphics[width=8cm]{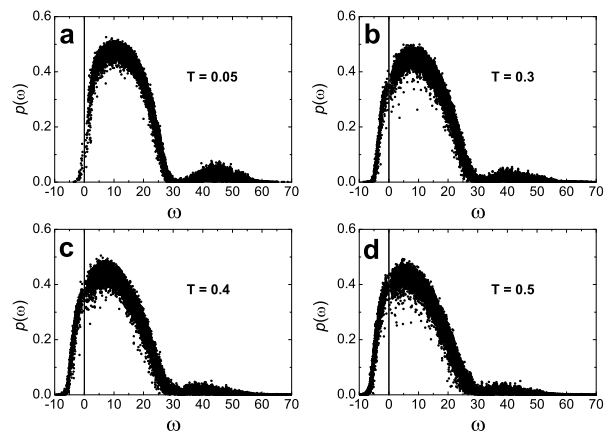}
\caption{Participation ratio at different temperatures. (a) $T=0.05$; (b) $T=0.3$; (c) $T=0.4$; (d) $T=0.5$. The imaginary frequencies are shown as negative frequencies. $N=50, M=100$.}
\label{pRatioFull}
\end{figure}

On the microscopic scale, the viscoelastic properties are controlled by the vibrational density of states $D(\omega)$ and affine force field $\Gamma(\omega)$ (see Eq. \ref{CModulus}), both of which contribute to non-affine softening. The density of vibrational states for a polymeric amorphous solid at low temperatures consists of two prominent features (Fig. 4(a)): a large peak upon normalizing by Debye's $\sim \omega^{2}$ law, the so-called `boson peak', which is associated with LJ interactions between beads; and a plateau at higher frequencies corresponding to FENE bond and collective bond-LJ vibrations \cite{dePablo2,macro2018}. The boson peak can be measured by neutron inelastic scattering experiments for instance for polyethylene \cite{PE_DOS} or PDMS \cite{PDMS2016}.  

An increase of temperature leads to the shift of boson peak and the FENE-bond plateau to lower frequencies (Fig. 4(a)), which increases the non-affine contribution (see Eq. \ref{CModulus}) and, thus, makes the material softer. The increase of the boson peak and its shift to lower frequencies clearly gives rise to an increase (in absolute value) of the non-affine integral due to the weight in the denominator which gives high weight to the low-$\omega$ part of the VDOS (Fig. 4(c)).

Clearly the VDOS in Fig. 4(a) has a model specific shape. Additional interactions could bring another peaks as we have shown earlier by considering the influence of angular potentials on the vibrational density of states\cite{macro2018}. In that case angular interactions gave additional contributions to the Hessian. The analogous procedure could be performed for dihedral interactions, hydrogen bonds etc. Generally, the characteristic energy of an interaction defines the part of the VDOS it enters\cite{macro2018}. Additional interactions will also affect the viscoelastic properties\cite{stiffness,hoy2018,Song2018}. The main basis of our approach is the harmonic approximation. If this assumption holds we expect that the method will be useful for study of all-atom models as well as coarse-grained models obtained from all-atom models by proper renormalisation of cohesive interaction strength as a function of temperature\cite{Song2018} or direct coarse-graining\cite{vur2017,vur2018}.

$\Gamma(\omega)$ also increases strongly in the range of $20\le\omega\le30$ (Fig. 4(b)). The large amplitude and fluctuations of $\Gamma(\omega)$  in the high frequency part does not play a significant role for non-affine contributions due to the fast growth of $1/\omega^2$ part in the integrand in Eq. \ref{CModulus}. Thus, both the density of INMs and the correlator of affine force field show features leading to the softening of the material with increasing $T$. It is clear that in our case the crossover from solid to liquid state above $T_g=0.4$ does not bring any new microscopic signatures of a transition, i.e. the microscopic INM-based non-affine approach, here quantitatively validated against simulations, corroborates the idea of gradual and continuous amorphous solid-liquid crossover.

\section{Conclusions}

Prediction of the dynamic mechanical properties of amorphous solids and liquids remains an open challenge in condensed matter physics. In this article, we have combined microscopic dynamical information about internal stresses in the form of INMs with an athermal non-affine deformation theory.
The combined approach is capable of describing the viscoelastic properties of polymer glasses,
achieving quantitative predictions of $G'$ across most frequencies and temperatures (except where the structural relaxation rate begins to compete with external shearing frequency) and qualitative agreement of $G''$ above and below $T_g$.
The drop of the low-frequency storage modulus at $T_g$ is linked to growth of the so-called boson peak (proliferation of soft modes above the Debye $\omega^{2}$ level) in the VDOS and its shift towards zero frequency/energy, and with the gradual appearance of non-propagating, unstable relaxation modes (imaginary eigenfrequencies) below $T_g$. All other microscopic features of the vibrational excitations and microscopic dynamics change continuously and gradually across $T_g$. Our results support the view that the glass transition represents a fundamental continuity between the liquid and the amorphous solid state \cite{kostyareview,kostyaPRL2017}, at least for its mechanical manifestation.

\appendix

\section{Simulation of mechanical spectroscopy}

\begin{figure*}\center
\includegraphics[width=16cm]{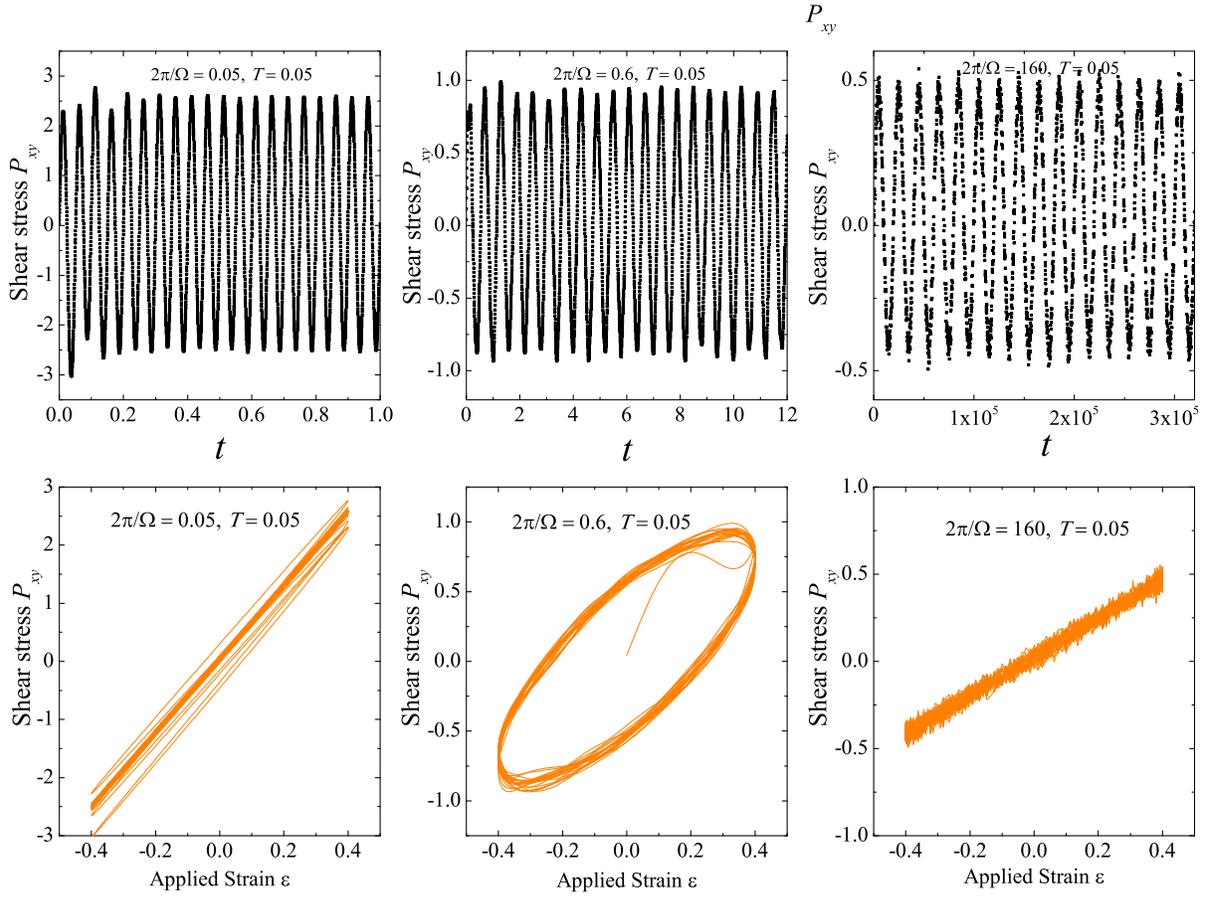}
\caption{Top: Instantaneous shear strain as a function of time for low, middle and high frequencies, correspondingly (left to right).  Bottom: Evolution of the instantaneous shear stress during cycles of applied shear strain for low, middle and high frequencies, correspondingly(left to right).}
\label{Lissajous}
\end{figure*}

In order to obtain $G',G''$ from molecular dynamics simulations we have used a built-in function in LAMMPS called wiggle. It is capable of simulating the shear strain and measuring the corresponding instantaneous stress $P_{xy}$. In Fig. \ref{Lissajous} we show an example of the data for the time dependence of $P_{xy}$ as well as the corresponding the stress-strain curves (Lissajous figures) for three different frequencies and $T=0.05$.

\section{Equivalence of results obtained from simple and pure shear in the limit of small deformations}

In simulations we use simple shear to obtain the viscoelastic moduli. The theory, however, predicts the values of these quantities for the pure shear deformation. Here we show that in the case of small deformations the results produced by a simple shear are equivalent to the results produced by a pure shear.

\begin{figure}\center
\includegraphics[width=8cm]{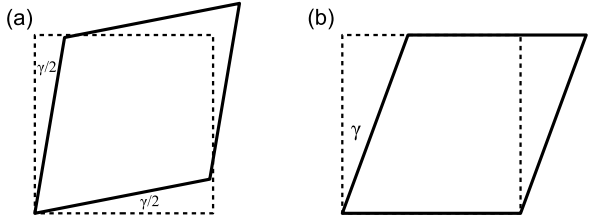}
\caption{Simple vs. pure shear deformations. (a) illustrates pure shear. The perimeter of the box is kept constant while the cross section area decreases with an increase of $\gamma$. (b) shows simple shear. The cross-section area stays constant while right and left side of cross section parallelogram becomes longer.}
\label{simplepure}
\end{figure}  

The following deformation tensors for the coordinate transform $\mathbf{r}'=\hat{u}\mathbf{r}$ for pure and simple shear, correspondingly, read:

\begin{equation}
\hat{u}_{\mathrm{pure}}= \left(\begin{matrix} \cos(\gamma/2) & \sin(\gamma/2) & 0 \\ \sin(\gamma/2) & \cos(\gamma/2) & 0 \\ 0 & 0 & 1\end{matrix}\right),
\end{equation}

\begin{equation}
\hat{u}_{\mathrm{simple}}= \begin{pmatrix} 1 & \tan(\gamma) & 0 \\ 0 & 1 & 0 \\ 0 & 0 & 1\end{pmatrix}.
\end{equation}

We now consider only central forces with $U(\gamma) = \frac{\kappa}{2} \sum_{ij}\left(r_{ij}(\gamma)-r_0\right)^2$. For the pure shear:  

\begin{eqnarray}\nonumber
&&r'_{ij}=\sqrt{r_{ij}^2+2r_{ij}^xr_{ij}^y\sin\gamma},\, \frac{\partial r'_{ij}}{\partial\gamma}\bigg\vert_{\gamma=0} = r_{ij}\hat n_{ij}^x\hat n_{ij}^y,\\
&&\frac{\partial r'_{ij}}{\partial \mathbf{r}_{i}}\bigg\vert_{\gamma=0} = -\hat n_{ij}.
\end{eqnarray}
Hence, the affine shear modulus and affine force field are (cf. definitions in the SI of \cite{stiffness}): 
\begin{equation}
G^A_{pure}=\kappa\sum_{ij}r^2_{ij}\left(\hat n_{ij}^x\hat n_{ij}^y\right)^2,\, \mathbf{\Xi}_{i,\mathrm{pure}}=-\kappa\sum_{ij}^Zr_{ij}\hat n_{ij}^x\hat n_{ij}^y\hat n_{ij}.
\end{equation}

For simple shear
\begin{eqnarray}\nonumber
&&r'_{ij}=\sqrt{r_{ij}^2+2r_{ij}^xr_{ij}^y\tan\gamma+(r_{ij}^y\tan\gamma)^2},\\\nonumber&& \frac{\partial r'_{ij}}{\partial\gamma}\bigg\vert_{\gamma=0} = r_{ij}\hat n_{ij}^x\hat n_{ij}^y,\\
&&\frac{\partial r'_{ij}}{\partial \mathbf{r}_{i}}\bigg\vert_{\gamma=0} = -\hat n_{ij}.
\end{eqnarray}

Since the derivatives of $r'_{ij}$ are the same as for pure shear, the values of the shear modulus and the affine force field should be the same as well. The reason for this equivalence is that for small deformations pure and simple shear can be converted one into the other by a solid rotation by an angle $\gamma/2$.

\section{Change of the number of nearest neighbours with temperature}

\begin{figure}
\includegraphics[width=8.7cm]{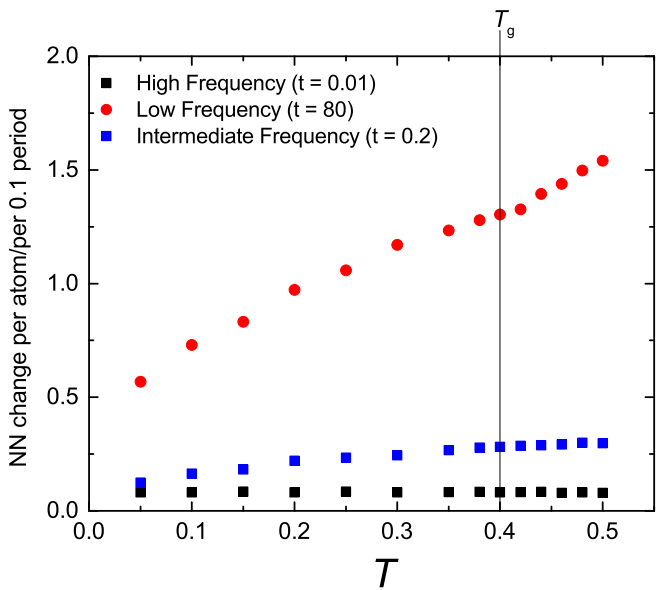}
\caption{Change of the number of nearest neighbors per atom as a function of temperature and frequency of the shearing force for periods $t=0.01$(black squares), $t=0.2$ (blue squares), $t=80$ (red circles). The positions of atoms are compared with the time resolution of $t/10$. The values obtained over 10 periods were averaged.}
\label{NNchange}
\end{figure}

In this section we show a small insight into the dynamical changes of the bead positions in order to illustrate the reasons for the applicability of the theory at non-zero temperatures. Two beads are considered to be neighbours if the distance between their positions is smaller than 1.2. Positions of the beads are compared at every tenth of an external force period. To enumerate the number of changes, count all new pairs which were formed, and all cases when two beads became separated, within this time. In Fig. \ref{NNchange} the average of change of the number of nearest neighbors is shown for three different frequencies and different temperatures. At all temperatures the beads move more in the case of low frequencies, i.e. at low frequencies the beads have time to adjust their positions and move between the cages. With an increase of temperature more and more beads become mobile, which should lead to the violation of harmonic approximation and, hence, our non-affine approach.  This explains why the combination of INM and non-affine theory performs rather poorly at low frequencies. On the contrary, at high frequencies the bead positions are considerably more stable. Hence for high frequencies the harmonic approximation holds, which explains why non-affine theory produces correct predictions of viscoelastic moduli values even at temperatures around and above glass transition.

\begin{figure}
\includegraphics[width=8.7cm]{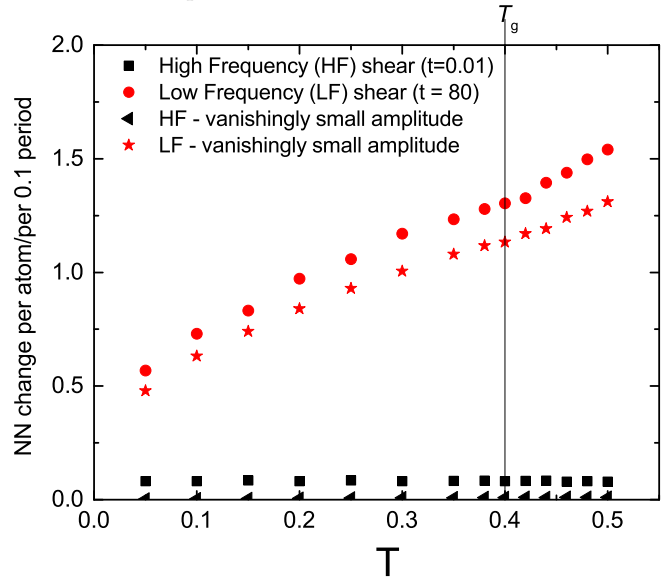}
\caption{The change of the neighbors in the case of sheared and undeformed samples at low and high frequencies. This plot illustrates that the absolute value of the difference is larger for the low frequency, which illustrates that at low frequencies many more atoms have time to relax their positions.}
\label{NNchange2}
\end{figure}

Definitely, the substantial part of a difference between the low and the high frequencies comes from the time resolution, which is period dependent. In Fig. \ref{NNchange2} we show the comparison of nearest neighbor number changes with and without shear. It is clear that the at low frequencies have the shear-induced changes have bigger contribution than at low frequencies.

\begin{acknowledgements}V.V.P. and A.Z. acknowledge financial support from the U.S. Army ITC-Atlantic and the U.S. Army Research Laboratory under cooperative Agreement No. W911NF-16-2-0091. C.N. acknowledges the Maudslay-Butler Research Fellowship at Pembroke College, Cambridge for financial support.
\end{acknowledgements}

\end{document}